\documentclass[aps,twocolumn,prd,epsf,showpacs,amsmath,amssymb]{revtex4}
\usepackage{dcolumn}
\usepackage{bm}
\newcommand{\e}{equation$\;$} 

\newcommand{\be}{\begin{equation}}
\newcommand{\ee}{\end{equation}}
\newcommand{\ba}{\begin{eqnarray}}
\newcommand{\ea}{\end{eqnarray}}
\newcommand{\ban}{\begin{eqnarray*}}
\newcommand{\ean}{\end{eqnarray*}}      
\newcommand{\n}[1]{\label{#1}}
\newcommand{\non}{\nonumber}
\newcommand{\eq}[1]{(\ref{#1})}

\newcommand{\ntr}{\ensuremath{{\nu(t,r)}}}
\newcommand{\str}{\ensuremath{\psi(t,r)}}
\newcommand{\ph}{\ensuremath{{\phi}}}
\newcommand{\dw}{\ensuremath{{d\Omega^2}}}

\newcommand{\E}{{\mathbb E}}

\begin{document}
\title{Regularity of initial data in dynamical massless scalar field models}
\author{Swastik Bhattacharya $^1$ and Pankaj S. Joshi $^1$}
\email{swastik@tifr.res.in, psj@tifr.res.in}
\affiliation{$^1$Tata Institute for Fundamental Research, Homi Bhabha Road,
Mumbai 400005, India}

\begin{abstract}

We discuss here the issue of regularity of initial data 
for dynamical spherically symmetric massless scalar field models
in a spacetime. Generalizing the known solutions of Einstein equations
given in this case by Wyman and Roberts, we examine the issue of 
regularity on a given spacelike surface, especially when the gradient 
of the field is spacelike. In particular, we isolate the class of models 
which would have necessarily a singularity at the center, and 
therefore these would be unsuitable for studying either gravitational 
collapse or dynamical evolutions in cosmology. 
\end{abstract}

\pacs{04.20.Dw, 04.20.Ex}
\maketitle

\section{Introduction}

Massless scalar fields coupled to gravity are of particular 
interest in both gravitational collapse situations as well as for
the cosmological scenarios. In cosmology, special 
importance is attached to the evolution of a scalar field, 
which has attracted a great deal of attention in past decades. 
This is because one would like to know the behaviour for 
fundamental matter fields towards understanding the transition 
from matter dominated regime to dark energy 
domination (see e.g. 
~\cite{nunes} 
and references therein). Scalar fields are of much 
interest in view of the inflationary scenarios that govern the 
early universe dynamics, because such a field can act 
as an `effective' cosmological constant in driving 
the inflation 
~\cite{nde}.

In gravitational collapse studies, the nature of singularity 
for the evolving massless scalar fields has been examined and 
a number of numerical and analytical works have been done 
in recent years, mainly on spherical collapse models 
~\cite{scalar1}-\cite{scalar8}.
The perspective here is mainly that of the cosmic censorship 
hypothesis, in the sense that one would like to understand the 
nature of the spacetime singularity developing in collapse in order
to know whether it is within a black hole or it could be a naked
singularity not covered within an event horizon, in violation
to the censorship conjecture.

Both the collapse and cosmological models, however,  
share one important physical feature, namely that except for 
the possible initial (in the case of cosmology ) and the final 
(can be applicable for both the cases of collapse and cosmology) 
singularities, the other epochs should be regular in these models. 
Therefore, for any physical model, we expect to have a regular 
initial data without any singularities from which there is a dynamical 
evolution. If any solution to Einstein equations fails to satisfy 
this criterion, then that would be unsuitable as a model for 
gravitational collapse or cosmological scenarios.

In this note, we examine this question in some detail, 
and in particular we isolate a rather wide class of massless scalar 
field models which are unsuitable as either collapse models or 
in cosmology, because they do not have a regular initial data 
and admit a singularity necessarily. We do not assume here
or discuss any particular solution, but a general class of models
is considered. Clearly, information such as this would be useful 
as one needs to avoid such classes where initial data 
is singular, for any physically realistic discussion. 
In what follows, we shall examine whether it is possible to 
have regular initial data for the general class of spherically symmetric 
massless scalar field models, where the gradient of the scalar field 
is spacelike on the initial spacelike hypersurface. The consideration 
would also hold as long as the gradient remains spacelike in the
neighborhood of the center, but could be timelike in other regions
of the initial surface.

The static limit of this class of models is the class of solutions 
given by Wyman 
\cite{Wyman}. 
The solutions given by Wyman are for 
spherically symmetric static models for a massless scalar 
field coupled to gravity, where the gradient of the scalar field 
is spacelike. We shall show that, in general, even for any dynamical
class of such models, if the matter is distributed in a ball around 
the center of spherical symmetry, then there is necessarily a 
singularity at the center. In other words, one cannot have a 
regular initial spacelike surface in this case, which is necessary 
for a physical collapse model.  Therefore this class of models 
is not suitable for studying collapse. As an example of the violation 
of regularity, we can consider the static solutions given by 
Wyman, all of which have a singularity at the center. Of course, 
these models, being static, cannot provide any insight on
gravitational collapse or dynamical evolutions. Unfortunately, 
not many dynamical solutions of the massless scalar field 
coupled to gravity are known. Among the existing few, the 
Roberts solution 
\cite{scalar4} \cite{scalar5} \cite{Oshiro} 
is the one which has been studied in some detail in literature. 
It has again a central singularity at all epochs. In this case, 
however, the gradient of the scalar field is timelike in some part 
and spacelike in other parts of the spacetime.

The existence of a central singularity in these models, 
when the gradient of the scalar field is spacelike, is in 
agreement with our work in this paper. In this sense, the result 
presented here may be considered as providing a generalization
of this important feature observed in these known classes 
of explicitly known solutions for the case of a massless 
scalar field. For the sake of clarity, we emphasize that we shall 
neither give here any explicit solutions nor outline any method 
for doing so in the present paper. Our aim is in fact to show 
that even when it is not possible to completely solve the 
Einstein equations, it is possible to know about certain 
important physical features of the solutions, such as the 
regularity of the models, as in the present case.

The plan of this paper is as follows. First, we set up the 
basic formalism by specifying the coordinate system which we 
work with. Then we briefly review the two classes of solutions 
given by Wyman and Roberts, and we note the presence of 
central curvature singularities in both these cases. In the 
next section, we then generalize this feature to demonstrate 
the breakdown of the regularity at the center, whenever the 
gradient of the scalar field is spacelike. The existence of central 
singularities in both the Wyman and Roberts solution are also 
discussed in view of this result. Finally, we consider the 
implications of this result in the context of the dynamical 
evolution of a massless scalar field coupled to gravity.

\section{Massless scalar fields}

In our analysis here, we consider a four-dimensional spacetime 
manifold which has spherical symmetry. The massless scalar field $\ph(x^a)$ on 
such a spacetime manifold $(M, g_{ab})$ is described by the Lagrangian, 
\be
{\cal L}=-\frac{1}{2}\ph_{;a}\ph_{;b}g^{ab}.
\n{lag}
\ee
The corresponding Euler-Lagrange equation is then given by, 
\be
\ph_{;ab}g^{ab}=0,
\n{el}
\ee
and the energy-momentum tensor for the massless scalar field, as calculated 
from the above Lagrangian, is given as 
\be
T_{ab}=\ph_{;a}\ph_{;b}-\frac{1}{2}g_{ab}\left(\ph_{;c}\ph_{;d}
g^{cd}\right).
\n{emt}
\ee
The massless scalar field is a {\it Type I} matter field
\cite{haw},  
{\it i.e.}, it admits one timelike and three spacelike 
eigen vectors. At each point $q\in M$, we can express the tensor 
$T^{ab}$ in terms of an orthonormal basis $(\E_0,\E_1,\E_2,\E_3)$, 
where $\E_0$ is a timelike eigenvector with the eigenvalue $\rho$.
The vectors $\E_{\alpha}$ $(\alpha=1,2,3)$ are three spacelike eigenvectors 
with eigenvalues $p_\alpha$. The eigenvalue $\rho$ represents 
the energy density of the scalar field as measured by an observer 
whose world line at $q$ has an unit tangent vector $\E_0$, and 
the eigenvalues $p_\alpha$ represent the principal pressures 
in three spacelike directions $\E_\alpha$.

We now choose the spherically symmetric coordinates 
$(t,r,\theta,\phi)$ along the eigenvectors $(\E_0,\E_\alpha)$. 
Such a reference frame describes {\it comoving} coordinates
which we use here. As discussed in 
\cite{landau}, 
the general spherically symmetric metric in comoving 
coordinates can be written as,
\begin{equation}
ds^2= e^{2\ntr}dt^2-e^{2\str}dr^2-R^2(t,r)\dw,
\label{metric}
\end{equation}
where $\dw$ is the metric on a unit two-sphere, and 
we have used the two gauge freedoms of two variables, namely, 
$t'=f(t,r)$ and $r'=g(t,r)$, to make the $g_{tr}$ term in the 
metric and the radial velocity of the matter field to vanish. 
That means that the energy-momentum tensor is diagonal
in such a coordinate system, and therefore the eigen values
$\rho$ and $p_{\alpha}$s have definite physical meaning of
mass-energy density and pressures. We note that we still 
have two scaling freedoms of one variable available here, 
namely $t \to f(t)$ and $r \to g(r)$. We note that the metric
function $R$ here denotes the physical radius of the 
matter cloud.

As we are considering here spherically symmetric spacetimes, 
we have $\ph=\ph(t,r)$ necessarily.
Furthermore, from  \e\eq{emt}, we can easily see that in the 
comoving reference frame with the metric given by \eq{metric},
$T_{10}=\phi' \dot{\phi}$. It follows therefore that
we must have necessarily 
$\ph(t,r)=\ph(t)$ or $\phi(t,r)=\phi(r)$, 
because the energy-momentum tensor here is necessarily 
diagonal.

As noted above, we have adopted here the comoving 
coordinates. This coordinate system had been used earlier to 
discuss the massless scalar fields, by Wyman 
\cite{Wyman}, 
Xanthopoulos and Xanias 
\cite{Xan}, and others. Some of these authors 
used this coordinate system to study the static case,
and also some dynamical cases were examined. 
However, it is also the case that this coordinate system 
is not used at times to study the dynamical evolution of 
a massless scalar field coupled to gravity. The main reason 
for that is, the comoving coordinate system breaks down 
if and when the gradient of the scalar field becomes null 
as the field evolves. However, this does not pose any problem 
as far as our purpose is concerned, because what we want 
to show here is that, for a massless scalar field with a spacelike 
gradient, there cannot exist any regular spacelike hypersurface 
from which to start the dynamical evolution of the gravitational 
collapse in a regular or non-singular manner. What we examine 
below is to consider the scalar field configuration only at an 
epoch of constant time, when the gradient of the field 
is spacelike.

\section{The Wyman and Roberts solutions}

We first discuss here the massless scalar field 
solutions given by Wyman 
\cite{Wyman}, 
and Roberts 
\cite{scalar4} \cite{scalar5} \cite{Oshiro}, 
respectively.
As mentioned above, Wyman used comoving coordinates 
in his analysis of the static massless scalar field spacetimes. 
Then there are two possible cases that arise, namely, 
$\phi(t,r)=\phi(t)$ or $\phi(t,r)=\phi(r)$. 
In the first case, the gradient of the scalar field is timelike 
everywhere, whereas it is spacelike in the latter case. 
For the first case, the Einstein equations are difficult to solve 
and can be solved only for a special case.The Einstein 
equations can be solved, however, when the gradient of the 
scalar field is spacelike, and a general solution can be 
obtained in this case. In this second case, the line element 
given by Wyman is 
\begin{equation}
\begin{split}
 ds^2=&  e^{\alpha r^{-1}}dt^2 - e^{-\alpha r^{-1}}[\frac{\gamma r^{-1}}{\sinh{\gamma r^{-1}}}]^4 dr^2 \\ -&
  e^{-\alpha r^{-1}}[\frac{\gamma r^{-1}}{\sinh{\gamma r^{-1}}}]^2 r^2 d\Omega^2 \label{Wyman}
\end{split}
\end{equation}

This generalises most of the previous static solutions 
obtained for a massless scalar field, which were given by 
various authors such as
Buchdal \cite{Buchdal} and 
Yilmaz \cite{Yilmaz}. 
Newman, Janis, and Winicour 
\cite{JNW} 
also discussed this solution in the comoving 
coordinate system. But they used a scaling of the 
coordinate radius $r$ which was different from the one 
that was used by Wyman. 

In this class of models given by Wyman, there is 
{\it always} a strong curvature singularity present at the center 
$R=0$ at all epochs. The Ricci scalar diverges at the singularity 
at $R=0$. From the line element \eqref{Wyman}, we see that
the metric component $g_{00}$ diverges at the center 
if $\alpha>0$ but is finite for $\alpha=0$.

The dynamical solution given by Roberts 
\cite{scalar4}, Brady \cite{scalar5} 
and Oshiro et al 
\cite{Oshiro} 
has the following form when written in double null 
coordinates,
\ba
 ds^2&=& - du dv + R(u,v)^2 d\Omega^2 \non\\
R(u,v)&=& \frac{1}{2} \sqrt{[(1-p^2)v^2-2vu+u^2]}\;.
\ea
The Ricci scalar here takes the form 
\begin{equation}
 R_c= \frac{p^2uv}{2R^4}
\end{equation}

There are three qualitatively different subclasses of 
this solution. For all the three subclasses, however, a singularity 
is always present at the center of the coordinates, where 
the Ricci scalar diverges. For the subclass $0<p<1$, the 
gradient of the scalar field is first timelike, and then it changes 
over to a spacelike vector. These two different regions are 
separated by a null hypersurface. In both the regions, 
one can set up different comoving coordinate systems,
which however, would break down at the null hypersurface joining 
them. In the region where the gradient of the scalar field is 
spacelike, the spacetime metric in the comoving coordinates 
has the form, 
\begin{equation}
\begin{split}
ds^2=& dt^2- \frac{1}{4} \frac{(1-p)^2[(1-p^2)r+2p]^2}{p^2[(1-p)r+1]^2} t^2 dr^2\\ -&
 \frac{(1-p)}{4} \frac{r[(1-p^2)r+2p]}{[(1-p)r+1]}t^2 d\Omega^2
\end{split}
\end{equation}
For the other two subclasses of the solution, the gradient 
of the field is always timelike. Because of the ever-present spacetime
singularity at the center $r=0$, it is not possible to find a regular initial 
spacelike hypersurface in this model where one can define the 
regular initial data from which to develop and evolve the 
gravitational collapse of the massless scalar field.

\section{The dynamical models with a spacelike gradient for the scalar field}

In this section, we consider the issue of 
the regularity for the dynamical models for which the gradient 
of the massless scalar field is spacelike during the evolution. 
We note that, even if the evolution of the field began
in such a manner that the gradient of the scalar field is
spacelike everywhere on the given initial spacelike surface,
it is possible that later in the evolution it may change to 
being timelike, either everywhere on a given later spacelike
surface, or in some regions of the same. The conclusions
given below remain valid as long as the gradient is spacelike
in some neighbourhood of the center. What we show 
below is, as long as a neighbourhood of the center has
a spacelike gradient for the scalar field, the center must
admit a curvature singularity, and therefore an evolution 
from a regular initial data is not possible.

To do this, we first write down the Einstein equations 
in the comoving coordinates. For the metric (\ref{metric}),
and using the following definitions,
\begin{equation}
G(t,r)=e^{-2\psi}(R^{\prime})^{2},\;\; H(t,r)=e^{-2\nu} (\dot{R})^{2}\;,
\n{eq:ein5}
\end{equation}
\begin{equation}
F=R(1-G+H)\;,
\label{eq:ein4}
\end{equation}
we can write the independent Einstein equations for the 
spherical massless scalar field (in the units $8\pi G=c=1$) 
as below (see \cite{GJ1}),
\begin{equation}
 \rho=\frac{F'}{R^2 R'},
\end{equation}
\begin{equation}
 P_r=-\frac{\dot{F}}{R^2 \dot{R}},
\end{equation}
\begin{equation}
 \nu'(\rho+P_r)=2(P_\theta-P_r)\frac{R'}{R}-P_r'\label{e3},
\end{equation}
\begin{equation}
 -2\dot{R}'+R'\frac{\dot{G}}{G}+\dot{R} \frac{H'}{H}=0\label{e1}, 
\end{equation}
In the above, the function $F(t,r)$ has the interpretation 
of the mass function for the matter field, in that it represents 
the total mass contained within a coordinate radius $r$.

For the class of models, $\phi=\phi(r)$, the components of the 
energy-momentum tensor are given by,
\begin{equation}
T^t_t=-T^r_r=T^{\theta}_{\theta}=T^{\phi}_{\phi}=\frac{1}{2}e^{-2\psi}\ph'^2
\end{equation}
It follows that the equation of state in this case, 
which relates the scalar field energy density and pressures 
is then given by,
\begin{equation}
 \rho=P_r=-P_{\theta}
\end{equation}
Putting the expressions of density and pressure in the Einstein equations, 
we have,
\begin{equation}
\frac{1}{2}e^{-2\psi}\phi'^2=\frac{F'}{R^2R'}=-\frac{\dot{F}}{R^2\dot{R}}
\end{equation}
We can also write from \eqref{e3},
\begin{equation}
\phi''=(\psi'-2\frac{R'}{R}-\nu')\phi' \label{phi'}
\end{equation}
and the equations \eqref{eq:ein4}, \eqref{e1} and  \eqref{phi'} 
can be integrated once to give,
\begin{equation}
\phi'= \frac{e^{\psi-\nu+b(t)}}{R^2} \label{phi'1},
\end{equation}
where $b(t)$ is an arbitrary function of the time coordinate. 
The Ricci scalar $R_c=g^{\mu\nu}R_{\mu\nu}$ can be written as 
$R_c=-2e^{-2\psi}\phi'^2$. Using \eqref{phi'1} we now get, 
\begin{equation}
R_c=-\frac{e^{-2\nu+b(t)}}{R^4}. \label{sing}
\end{equation}

In a physically reasonable collapse model, the initial 
spacelike surface should be regular. This implies that $g_{\mu \nu}$ 
and $g^{\mu \nu}$ are finite and regular on that surface. 
From \eqref{sing}, it is clear that on that surface, at $R=0$, 
there will be a singularity. Unless we are considering the 
collapse of a thick shell of scalar field, the physical radius $R$ 
of the spherical distribution of scalar field must be zero at the center. 
Therefore, there will always be a singularity at $R=0$ during the 
time evolution of a scalar field of the form $\phi=\phi(r)$, unless 
$e^{-2\nu}$ goes to zero at the center $R=0$ at least as fast 
as $R^4$. This later situation implies a breakdown of the coordinate 
system at the center $R=0$. For the static solutions given by 
Wyman, as $R$ goes to zero, $e^{-2\nu}$ never goes to zero 
as fast as $R^4$. Therefore there is a singularity always present 
at the center for these models. In the dynamical solution given 
by Roberts, $e^{-2\nu}=1$ always, and therefore from 
\eqref{sing}, it follows that the Ricci scalar diverges at the 
center, $R=0$. 

It follows that our consideration here generalizes the 
behaviour of existence of a singularity at the center $R=0$, 
as seen in the Wyman and Roberts models, to the general 
class of all models where the gradient is spacelike in a 
neighbourhood of the center. It thus follows that this class 
cannot be used for any considerations related to collapse 
or cosmology, where dynamical evolution from a regular 
initial data is a necessary condition. 

We noted above that a breakdown of the coordinates 
at $R=0$ could be an alternative to singularity. Let us now 
consider what such a breakdown of the comoving coordinate 
system at $R=0$ would signify in the case, if there is 
no singularity. For $R_c$ to be finite at $R=0$, 
$e^{-2\nu}\sim R^4$ as $R$ goes to zero. If $\phi,_\mu$ is 
spacelike, then a comoving shell with non-zero physical 
radius $R$ cannot evolve in such a way so that the physical 
radius becomes zero without hitting the singularity. In that 
case, we can consider the center as a comoving world line. 
Therefore, as one goes towards the center, along the 
comoving world lines $r=const.$ near the center, the proper 
time must necessarily diverge. In other words, the length 
of the comoving world lines diverges as one approaches the 
center. While this may not be as serious as having a singularity 
at the center, it does imply some kind of violation of regularity 
at the center. In such a case, this may not mean a violation 
of the regularity of the initial data, but rather that the dynamical 
evolution is not regular in some sense. The collapse
or the evolution at the center then never reaches a singularity
in future for an infinite proper length and the collapse is 
`freezing' in the sense above.

\section{Discussion}

It is seen that for the class of models considered here, 
the center is not regular at any epoch. In fact, this feature is seen 
in the static spacetimes described by Wyman's solution also. All 
such static solutions with $\phi=\phi(r)$ are singular at $R=0$. It 
follows that the class of massless scalar field models with 
$\phi=\phi(r)$ is thus not adequate for the investigation of 
dynamical gravitational collapse evolution of the field from a regular 
initial data 
\cite{GJ1}. 
Therefore, whenever we are interested in such 
time-dependent models, or key issues related to the 
gravitational collapse phenomena, we must focus on the 
$\phi=\phi(t)$ class of massless scalar fields only.

Another important point to note here is that the central 
singularity could also exist even if $\phi,_\mu$ is spacelike at a 
neighbourhood of the center. This is because in that case, we 
could set up a comoving coordinate system which covers this 
neighbourhood during the dynamical evolution. This implies 
that in that patch, the Einstein equations can be written in the same 
form as given in this paper. Then it follows that at the center, 
there would be a violation of regularity. This means that one can 
extend the result to the cases where the gradient of the scalar 
field is spacelike at the center. In those cases also, it is not possible 
to have regular initial data, from which the collapse would evolve. 

Another interesting outcome of this result is that during the 
evolution, if $\phi,_\mu$ changes to a spacelike vector from a 
timelike one, the center of the scalar field cloud would 
cease to be a regular point. It would also be interesting to ask whether
the singularity in question is naked or covered within an event
horizon. While the answer can be found in some special cases, 
we do not address this issue here presently, which would
in general require more information on the nature of the
solutions for the massless scalar field case.


\begin{thebibliography}{99}
 
\bibitem{nunes} N. J. Nunes, J. E. Lidsey, Phys.Rev. D 
{\bf 69} (2004) 123511.

\bibitem{nde}Linde A.D., {\it Particle physics and inflationary cosmology},
Harwood Academic, 1990.

\bibitem{scalar1}D. Christodoulou, Ann. Math.{\bf 140}, p.607 (1994).

\bibitem{scalar2} M. W. Choptuik, Phys. Rev. Lett. {\bf 70}, p9 (1993). 
 
\bibitem{scalar3}C. R. Evans and J. S. Coleman, Phys. Rev. Lett. 
{\bf 72}, p.1782 (1994). 

\bibitem{scalar4}M. D. Roberts, Gen. Relat. Grav. {\bf 21}, p.907 (1989). 
 
\bibitem{scalar5}P. R. Brady, Class. Quant. Grav. {\bf 11}, p.1255 (1995). 

\bibitem{Oshiro} Y. Oshiro, K. Nakamura and A. Tomimatsu, Prog. Theor. 
Phys., {\bf 91}, p.1265 (1994).
 
\bibitem{scalar6}C. Gundlach, Phys. Rev. Lett. {\bf 75}, p.3214 (1995). 
\bibitem{scalar7}E. Malec, Class.Quant.Grav. {\bf 13} p. 1849 (1996).
\bibitem{scalar8}R. Giambo, F. Giannoni, G. Magli, {\it gr-qc/0802.0992,
gr-qc/0802.0157}

\bibitem{Wyman}Max Wyman; Phys Rev D, Vol 24, No 4, 1981

\bibitem{Xan}B.C.Xanthopoulos and T.Zannias: Phys. Rev.D, Vol 40, No 8, 1989

\bibitem{haw}S. W. Hawking and G. F. R. Ellis, {\it The large scale
structure of space-time}, Cambridge University Press, Cambridge (1973).

\bibitem{landau} Landau \& Lifshitz, {\it Classical theory of fields}, 
p.304 (1975).

\bibitem{Buchdal}H.A.Buchdal: Phys. Rev. 115,1325 (1959).

\bibitem{Yilmaz}H.Yilmaz:Phys. Rev. 111, 1417 (1958)

\bibitem{JNW} A.I.Janis, E.T.Newman and J.Winicour; Phys.Rev.Lett. 
{\bf 20}, p.878 (1968).

\bibitem{GJ1} P.S. Joshi, I.H. Dwivedi, Class. Quant. Grav.
{\bf 16}, p.41 (1999); R. Goswami and P. S. Joshi, Phys. Rev. {\bf D76}, 
p.084026 (2007).

\end{thebibliography}
\end{document}